\documentstyle[12pt,epsf]{article}
 \newcommand{\be}[1]{\begin{equation}\label{#1}}
\newcommand{\ee}{\end{equation}}

\begin{document} 
 \title{\bf Search for an unitary mortality law through a theoretical model\\
for  biological ageing}
 \author{A. Racco, M. Argollo de Menezes and T.J.P.Penna\\
 {\small e-mail address: adriana@if.uff.br, marcio@if.uff.br, tjpp@if.uff.br}\\ 
{\small Instituto de F\'{\i}sica, Universidade Federal Fluminense,}\\ {\small Av. Litor\^anea 
s/n, 24210-340, Niter\'oi, RJ, Brazil} } 
 \maketitle 
\begin{abstract} In this 
work we check the occurrence of the Azbel assumption of mortality within the framework of
 a bit string model for biological ageing. We reproduced the observed feature of
 linear correspondence between the fitting parameters of the death rate as obtained by
 Azbel with demographic data. \end{abstract}

 \vspace{8mm}

\section{Introduction} \par Since the pioneer work of Gompertz of 1825 on the exponential
increase of mortality with human age,
 there have been various attempts to quantify this relation between mortality rate and the ageing
process.  But there is no consensus on whether genetically programmed death exists or not. Some recent
works raise questions about the existence of a single mortality pattern, also discussing whether senescence
can be characterized by the increase of mortality with age. This features have been detected in the
analysis of medflies, where the mortality rate of the oldest medflies decreased with age  ~\cite{carey}, and in
 drosophilas, where it levels off ~\cite{curtsinger}~\cite{charles}.  There are also some questions about
 the existence of an absolute life expectancy for particular species. 
\par Recently, Azbel ~\cite{azbel} proposed an unitary mortality law for species based
on the analysis of demographic  and medfly data. His assertion is that this result is due only 
to genetic features. Based on this, we try to verify  if this assumption supported by the 
mutation accumulation theory of biological ageing, simulated on the computer with the bit 
string model ~\cite{penna}. \par The bit string model has been successfully applied
in many ageing problems, like the death rate of German  population ~\cite{zeits}, catastrophic 
senescence of pacific salmon ~\cite{salmon}, vanishing of northern  codfish ~\cite{cod}, 
longevity of trees ~\cite{trees}, advantages of sexual reproduction ~\cite{rita}~\cite{wolf}
 and others which reinforce the mutation accumulation theory. \par This paper is organized as
follows: in section $1$ we review the Azbel proposal ~\cite{azbel} and the different
  results found for homogeneous and heterogeneous populations. In section $2$ we review the bit
string model and  discuss the meaning of homogeneous and heterogeneous populations in this 
computer model. Finally, in section $3$ we   present the results and discussions. 

\section{The Azbel proposal} \par In a recent paper, Azbel ~\cite{azbel} proposed the existence
of a mortality law that could successfully explain the behaviour of different death rates like
human and medfly mortality curves, like the German death rate shown in fig.$1$. 
 His basic assumption is the existence of at least one Gompertz region in the death rate, which
means that it must exist some region in the mortality curve where the behaviour is predominantly
exponential. 

\begin{figure}[htbp] 
\epsfysize=2.7truein 
\centerline{\epsfbox{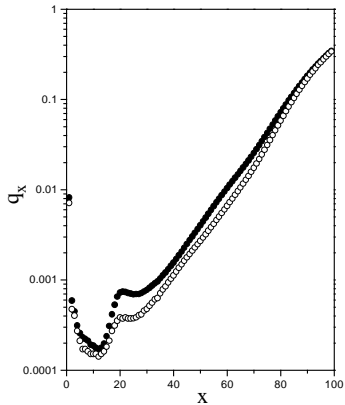}} 
\caption{Male $(\bullet)$ and female $(\circ)$ German death rates evidencing the Gompertz 
region.} 
\end{figure}

\par The mortality $q$ in a certain age $x$ at time $t$ is 
\be{eq1} q_x(t)=-\frac{N_{x+1}(t+1)-N_x(t)}{N_x(t)} \ee 
where $N_x(t)$ is the number of individuals alive
with age $x$ at time $t$. Considering time as a continuous variable, we have 
\be{eq2} q_x(t)=-\frac{d\ln N_x}{dt} .\ee 
Assuming that $q_x$ is an exponential function of time and integrating the above equation from 
$t-1$ to $t$ we get the better  definition of mortality rate: 
\be{eq3} q_{x}(t)=\ln{N_{x}/N_{x+1}}. \ee 
We can also write $N_{x+1}(t+1)=N_x(t)-D_x(t)$, where $D_x(t)$ is the number of deaths in time $t$ 
with age $x$, and obtain 
\be{eq4} q_x=- \ln(1- \frac{D_x(t)}{N_x(t)}). \ee 
Following the assertion that every death rate has its Gompertz region, where the logarithm of 
mortality rate is close to its linear regression
 \be{eq5} \ln q_x=a^{\ast}+bx .\ee 
\par Fitting the mortality curves of demographic data, Azbel determined the values of 
$a^{\ast}$ and $b$. In fact, Azbel worked with the dimensionless death rate 
\be{eq6} \ln \frac{q_x}{b}=a+bx \ee
 with $a=a^{\ast}-\ln b$. From each mortality curve it is obtained a pair $(a,b)$ and by
extensive study of demographic data of Japanese and Sweden population (which have low premature 
mortality), Azbel showed that the points determined by $a$ and $b$ follow a straight line and so 
can be related by $a=\ln A-b X$. With this we can write the death rate as \be{eq7} q_x=Ab 
\exp[b(x-X)]. \ee The striking feature is the behaviour of the parameters with respect to 
distinct populations (at different times): Japanese (1891-1990),
 Swedish (1970-1991) and German (1987) lifetables, the fittings give $X = 103 \pm 2$ years, $\ln
A = 2.4 \pm 0.2$. Fifty abridged Swedish (1780-1900) lifetables yield $X = 101 \pm 1$ year, $\ln
A = 2.4 \pm 0.1$; 62 abridged world-wide lifetables provide $X = 100.4 \pm 2.3$; $\ln A = 2.1
\pm0.3$, and all data combined (for a few billion males and females in different countries with
reliable statistics, over the period of two centuries) yield almost the same age $X=103\pm1$ and
number $\ln A=2.5\pm0.1$ . With all this data, Azbel proposes that as $X$ and $A$ are the same
within one species, they must be genetic factors. Since for these developed countries there is low
 premature mortality, we can say that the mortality in these countries is due principally to genetic factors. 
 Equation (7) is valid for sufficiently homogeneous cohorts, which means individuals with the
same genetic factors that will affect their life history. Homogeneous populations have smooth
death rates, since the life history of the individuals in normal situations is almost the same.
When the population is heterogeneous, then some features might change like the appearance of
arbitrarily old Methuselah's ( it has been observed that under the worst conditions, 30 of 3
million female medflies should survive a year, 18.6 times their life   expectancy at birth ~\cite{carey})
 and the linear increase of maximal lifespan with the initial population. Heterogeneous populations,
 as studied by Carey et al ~\cite{carey} with medflies, indicate the lack of exponential behaviour
 with \be{eq8} 1/N_x \propto (x-x_3)/N_0 \ee
 for $x$ sufficiently beyond $X$, where $x_3$ and $N_0$ are constants. If heterogeneity is
sufficiently high to allow $b \rightarrow 0$, in eq($7$) the main contribution to the region of
old ages of the death rate will come from individuals with smallest b. 

\section{The Bit-String Model and\\ Computer Simulations} \par In the bit-string
model~\cite{penna}, the age-dependent effects of ageing described by an evolutionary theory
~\cite{partridge} are taken into account by assigning to each individual in the population a
computer word (string) of , in our case, 128 bits, which should be interpreted as a
transcription of the genetic code. Each bit of the word will represent the activity of the ageing
process in that age.  The ages in which the deleterious mutations will affect the individual will
have their bits in the computer word set to one, while the others are all zero . After an age
$R$ the individual is mature and can reproduce (in this case, asexually) generating $B$ offspring,
or one child with probability $B$ if $B<1$. It is known that the transcription of th e genetic
code from parent to child is not perfect, and that the errors in the transcription accumulate.
These errors can be deleterious or even good to the individual, but the rate of deleterious
mutations is much higher then the rate of the good ones. I n the bit string model , when an
individual reproduces, its offspring will have the same genetic code of its parent, except for
$M$ new mutations randomly disposed on the string at the moment of birth. At the age that the
number of mutations in the gene tic code exceeds a certain threshold $T$, the individual dies.
Limitations of food and space are introduced into the model through a logistic factor, known as
Verhulst factor. This allow us to control the population size, which is of vital importance to
 do the computer simulations. After all individuals of the population have been tested, we count
one time step (in this paper, one time step corresponds to one year). \par Even though we have a
Verhulst factor to control the population size, we will count in the simulations the deaths due
only to genetic causes (fig.$2$) .
\begin{figure}[htbp]
\epsfysize=2.7truein
\centerline{\epsfbox{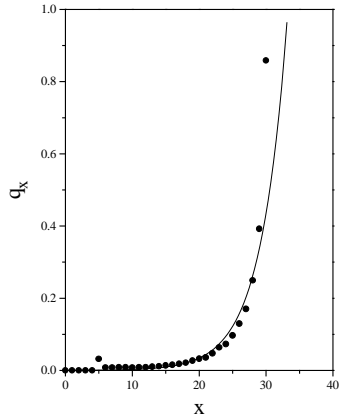}} 
\caption{Typical death rate obtained with the bit-string model
$(\bullet)$ with parameters $T=3,B=0.5,M=1.0$ and $R=12$. The solid line is the theoretical value
of death rate. } 
\end{figure}

\par Starting from a random population (i.e., individuals with different genetic code) of 10000
individuals, we wait a transient period until the size of population reaches an equilibrium state
and collect the informations of the system. 
 Setting all individuals of the population with the same intrinsic parameters guarantees us
sufficiently homogeneous cohorts when equilibrium (steady state) is reached and prevents us from
getting extremely fluctuating death rates due to the variability  of the lifespan introduced by
 varying the parameters (see ~\cite{penna} - ~\cite{trees}).  \par
In order to calculate the death rate of an homogeneous population, shown in fig.2, we took the
last 10000 steps of a total of 100000 (150000 for mutation rates $<1$) ~\cite{zeits}. 

\section{Results and discussions} \par We performed several tests controlling the parameters of
the bit-string model, first we just varied the seed for the random number sequence with all the
other parameters fixed. With this, we generate initial populations with different genetic code,
 which allow us to obtain different mortality curves. These random numbers also control which
individual will die under the pressure of the Verhulst factor. With this, we try to reproduce the
same species in different situations, as human population in different countries. When can also
calculate the death rate at different time steps of a single simulation, as did Azbel for
demographic data. 
 \par With minimum reproduction age $R=12$ all fittings of the Gompertz
 region were made for ages $>15$, and for each fitting we store the $(a,b)$ pair. Plotting this
pair in a linear scale we get the same qualitative result presented by Azbel, as seen in figure
$3$. Also, it can be observed that different mutation rates give us different straight lines of
the pair $(a,b)$, thus giving a suggestion of different species in the bit-string model
(fig.$3$). 
\begin{figure}[htbp] 
\epsfysize=2.7truein 
\centerline{\epsfbox{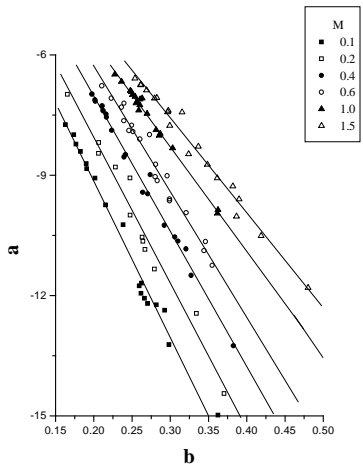}}
\caption{Linear behaviour of the death rate parameters $(a,b)$ obtained for different mutation
rates $M$. We have used $T=3,B=0.5,R=12$. Each point in the plot corresponds to a simulation with
a different seed.} 
\end{figure}
   
Observing the behaviour of the linear fittings obtained for different mutation rates in the
simulation and using the relation   between $M$ and $X$ obtained in the figure $4$, we can make 
an extrapolation of the results obtained by Azbel in order to have  an idea of how low have to be
 the mutation rate of the genetic code of human species. In order to have $X=103$ years
 ~\cite{azbel} we have $M\approx 0.001$, which is in agreement with the mutation rate proposed to
fit the female mortality in Germany ~\cite{zeits}. In order to describe effects like premature
mortality it is necessary to define other genetic factors in the  model, which is out of the scope
 of this work. \par Analyzing the behaviour of the parameters involved on the phenomenological law
 we can see that the characteristic age $X$   increases with decreasing mutation rate $M$ (fig.$4$),
 and that $b$ increases with increasing $M$(fig.$5$). 
 We can observe at the simulations that the characteristic age, $X$, is not the maximum lifespan,
 as already concluded in ref. ~\cite{azbel}. We can also check from our simulations that it is
smaller
 than the maximum lifespan. 
\begin{figure}[htbp] 
\epsfysize=2.7truein
\centerline{\epsfbox{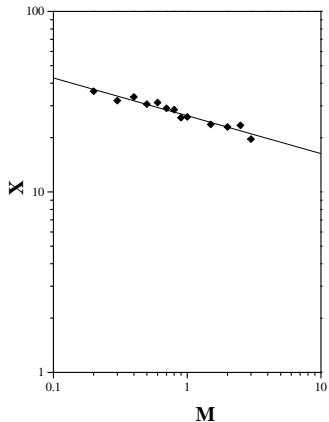}} 
\caption{Power law behaviour of $X$ with respect to $M$. The
solid line corresponds to the fitting $\log_{10} X=1.42-0.21~\log_{10} M$. The parameters $T,R$ and $B$ are
the same from figure $3$. } 
\end{figure} 

\begin{figure}[htbp] 
\epsfysize=2.7truein
\centerline{\epsfbox{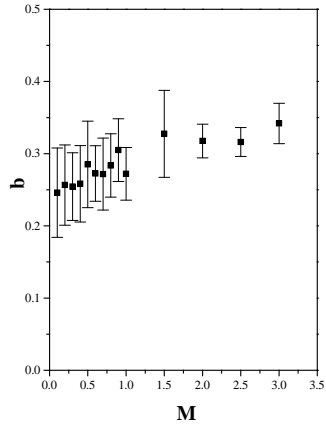}} 
\caption{Typical behaviour of $b$ for various mutation rates $M$
with $R,T$ and $B$ fixed with the same values of figure $3$.} 
\end{figure} 
\par With the parameters of the problem established, we can get back to the figure $2$, where we 
have already plotted the death rate obtained from the simulations $(\bullet)$ and the respective 
theoretical mortality law (line) obtained from eq$(7)$. 

\par In the treatment of heterogeneous populations we have to interpret first what is
heterogeneity in the bit-string model.  When we start the simulations we give different genotypes
 to the individuals, and these individuals compete with themselves and only the individuals with best
 genotypes survive and form the population of the steady state. More precisely, the genotypes disappear
 as a power law in time ~\cite{eva}. These out of equilibrium properties have already been
 proposed as an explanation of the oldest old effect in drosophilas ~\cite{charles}. 
  Our proposal is that the heterogeneity should be essentially an out-of-equilibrium state, where
individuals with very different characteristics coexist. The degree of heterogeneity is related
with the distance from the equilibrium state, so in order to  study heterogeneous populations we 
study the properties of the model far from equilibrium. In that region, as shown in figure $6$,
 is evident the non-exponential behaviour of the distribution of the individuals with age, as
 in ref ~\cite{carey}.  

\begin{figure}[htbp]
\epsfysize=2.7truein 
\centerline{\epsfbox{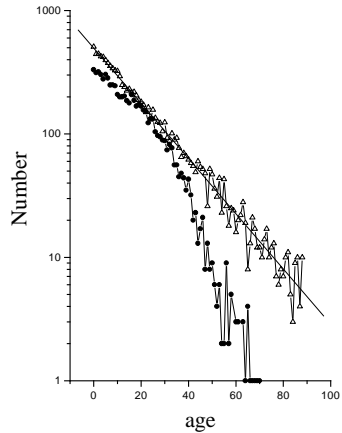}} 
\caption{Non-exponential behaviour of the
out of equilibrium population $(\bullet)$ in contrast to the equilibrium state $(\triangle)$.}
\end{figure} 
\par In summary, by simulating the bit-string model we succeeded in reproducing the
behaviour of the death rate parameters of both demographic and medfly data reinforcing the
mutation accumulation theory of biological ageing and the reliability of the bit-string model. 

\section{Acknowledgments} We thank D.Stauffer for fruitful discussions and suggestions, and
M.Azbel for providing copies of his work prior to publication. This work is partially supported
by Brazilian
 agencies CNPq and CAPES. 

 \end{document}